\documentclass[journal]{IEEEtran}

\usepackage{graphicx, subfigure}
\usepackage{amsmath}
\usepackage{cite}
\usepackage{amsfonts}
\usepackage{multirow}
\usepackage{url}
\usepackage{threeparttable}
\usepackage{rotating}
\ifCLASSINFOpdf
\else
\fi
\begin{document}
%
\title{Compression and Acceleration of Neural Networks for Communications}

\author{Jiajia~Guo,
        Jinghe~Wang,
        Chao-Kai~Wen, \IEEEmembership{\normalsize {Member,~IEEE}},
        Shi~Jin,~\IEEEmembership{\normalsize {Senior~Member,~IEEE}},
        ~and Geoffrey~Ye~Li, \IEEEmembership{\normalsize {Fellow,~IEEE}}
\thanks{J. Guo, J. Wang and S. Jin are with the National Mobile Communications Research Laboratory, Southeast University, Nanjing, 210096,
P. R. China (email: jjguo@mail.ustc.edu.cn, jwangeh@connect.ust.hk, jinshi@seu.edu.cn).}
\thanks{C.-K.~Wen is with the Institute of Communications Engineering, National Sun Yat-sen University, Kaohsiung 80424, Taiwan (e-mail: chaokai.wen@mail.nsysu.edu.tw).}
\thanks{G. Y. Li is with the School of Electrical and Computer Engineering, Georgia Institute of Technology, Atlanta, GA 30332 USA (e-mail: liye@ece.gatech.edu).}
}


\maketitle

\begin{abstract}
Deep learning (DL) has achieved great success in signal processing and communications and has become a promising technology for future wireless communications.
Existing works mainly focus on exploiting DL to improve the performance of communication systems.
However, the high memory requirement and computational complexity constitute a major hurdle for the practical deployment of DL-based communications. 
In this article, we investigate how to compress and accelerate the neural networks (NNs) in communication systems.
After introducing the deployment challenges for DL-based communication algorithms,
we discuss some representative NN compression and acceleration techniques.
Afterwards, two case studies for multiple-input-multiple-output (MIMO) communications, including DL-based channel state information feedback and signal detection, are presented to show the feasibility and potential of these techniques.
We finally identify some challenges on NN compression and acceleration in DL-based communications and provide a guideline for subsequent research.
\end{abstract}

%
\IEEEpeerreviewmaketitle

\section{Introduction}
%
%
%
%
\IEEEPARstart{I}{n} recent years, deep learning (DL) has brought many breakthroughs in various fields, such as computer vision and natural language processing.
Inspired by these successful applications, DL-based methods have gained a lot of attention from the communication community \cite{wang2017deep,8663966}.
Different from the traditional approaches that need rich expert knowledge, DL-based communication systems can automatedly discover the intricate structure from a large dataset.

The most existing literature explores the power of DL in wireless communications to improve the performance but seldom discusses the implementation challenges.
One of the most critical problems is the complexity of neural networks (NNs), including the large numbers of network weights and the high computational requirement.
Though NNs can be rapidly trained offline using powerful Graphics Processing Units (GPUs), the memory resources and computational units are limited at the real-time inference phase \cite{8114708}.
For example, in the fifth generation cellular systems (5G), the end-to-end latency should be within no more than 1 ms, so the DL-based algorithms should finish the inference of all NN-based modules in much less than 1 ms.
It is impractical for user equipment (UE) with limited resources (memory resources, computational units, and battery power) to realize the inference in such a short period.
CsiNet-LSTM in \cite{wang2018deep}, only compressing and reconstructing downlink channel state information (CSI), needs about 0.3 ms with a powerful NVIDIA 1080Ti GPU.
Obviously, these kind algorithms cannot be directly deployed  to practical communication systems.
Most existing DL-based algorithms are based on simulation using DL libraries, e.g., TensorFlow and PyTorch.
In these DL libraries, the NN weights are set as 32-bit floating point numbers by default, which not only occupies substantial storage space but also wastes precious computational resource.

Compared with the most traditional approaches, which store no weights and carry out limited iterations, the DL-based communication algorithms have to store up to millions of weights and need huge computational resources.
The computational complexity and memory requirement severely hinder the deployment of DL-based algorithms to communication systems and it becomes essential to design efficient and high-performance NNs for communication systems.
However, to the best of our knowledge, only few papers have taken the implementation of DL-based communication algorithms into consideration so far.

In this article, we first investigate the complexity trend of NN-based communications.
Then, we introduce the NN compression and acceleration techniques for communication systems.
Two case studies on DL-based CSI feedback \cite{wen2018deep} and signal detection\cite{samuel2019learning} in multiple-input-multiple-output (MIMO) communications are presented to show the feasibility and potential of the above techniques.
Finally, we highlight some open research issues of the NN compression and acceleration in communications.

The rest of this work is organized as follows. 
Section \ref{trend} explains the growing trend of NN complexity in DL-based communications.
In Section \ref{compression_techniques}, we introduce representative NN compression and acceleration techniques and their recent advances.
Then, two case studies, including the DL-based CSI feedback and signal detection, are prensented in Sections \ref{Two Case Studies}. 
Section \ref{conclusion} discusses the model compression and acceleration on communication algorithms and proposes several challenges.

\begin{figure*}[t]
    \centering 
    \includegraphics[scale=1.]{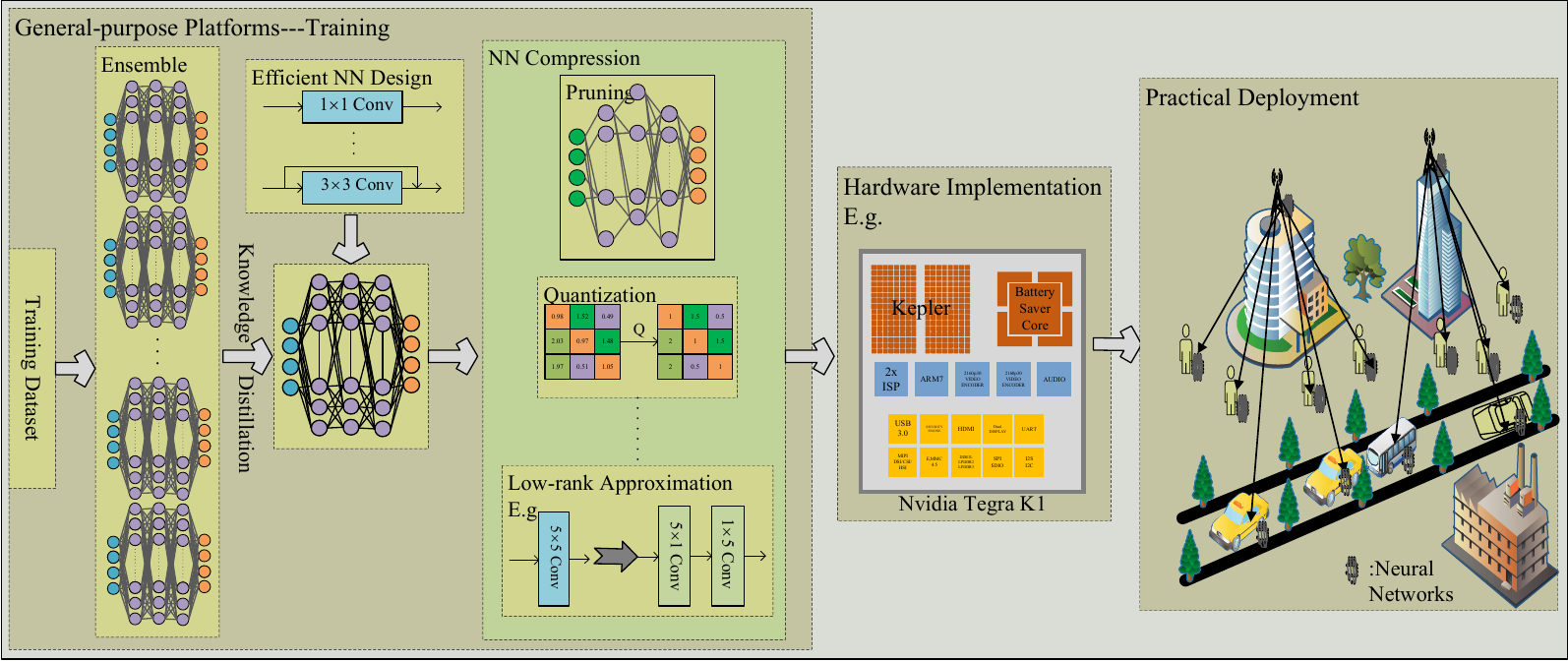}
    \caption{\label{overall}Overview of the training and implementation strategy of DL-based communication algorithms.}  
\end{figure*}

\section{Trend of Growing NN complexity}
\label{trend}
The NNs nowadays are pretty complicated compared with conventional communication algorithms.
Inevitably, the NNs will be more and more sophisticated in the future.

First, the current model-driven DL-based algorithms \cite{8715338} only implement the function of one or two modules in a communication system.
For example, only CSI feedback and signal detection are realized by CsiNet\cite{wen2018deep} and FullyCon\cite{samuel2019learning}, respectively.
But a complete communication system includes more modules, such as, source coding and decoding, channel coding and decoding, channel estimation, symbol detection, equalization, etc.
If NNs perform functions of all modules, the whole NN system will be much more complicated.

Furthermore, most of the existing DL-based communication algorithms focus on demonstrating what DL can bring to communications under simple scenarios.
For example, only limited literature takes the multiple-user scenario into consideration and most of the proposed NNs can only work under a certain channel model\cite{8054694}.
Future DL-based algorithms are expected to deal with much more complicated scenarios.
The requirement of memory space and computational resource will be further increased.

More and more novel techniques will be developed in future communication systems.
How to combine them with DL is an emerging problem.
Extra-large scale massive MIMO is a promising technology for the next-generation communication systems\cite{de2019non}.
More computational resource is required for signal processing in massive MIMO to obtain its benefits.
For instance, the fully connected (FC) layers for massive MIMO CSI feedback in \cite{guo2019convolutional} occupy over 95\% the number of the trainable weights, which is proportional to the square of the antenna number.
In brief, when DL meets extra-large scale massive MIMO, the NN weight number and complexity will drastically increase.

\section{NN compression and acceleration}
\label{compression_techniques}

We have witnessed a remarkable development in NNs, specifically in convolutional NNs (CNNs), across a wide range of areas. 
In order to achieve better performance and perform more functions, the scale of NNs is continuously expanding.
As a result, NNs are becoming model-complicated, memory-extensive, and computation-intensive. 
To tackle these issues, many approaches, including knowledge distillation, efficient NN design, pruning, quantization, and low-rank approximation, have been proposed over the past several years. 

As shown in Fig. \ref{overall}, we can address the issue in the DL-based communications using the following steps:
\begin{itemize}
\item[i)] the high-performance NNs are first trained without considering complexity.
\item[ii)] the dark knowledge achieved by these NNs and the efficient network design principles are utilized to design and train a compact NN model.
\item[iii)] the trained NNs are compressed by pruning, quantization, low-rank approximation, etc.
\item[iv)] the NNs are implemented on the task-specific hardware and deployed to practical environments.

\end{itemize}

In this section, we will provide some NN compression and acceleration techniques for communication algorithms.

\subsection{Knowledge Distillation}
Ensemble learning can improve the model performance by averaging the predictions from different models trained on the same dataset but at the expense of vast complexity increase.
The complexity of ensemble learning can be reduced by knowledge distillation, namely, a teacher-student network structure, which utilizes the dark knowledge achieved by the ensemble or cumbersome models (teacher) to train an efficient and compact model (student).
The student network can achieve a better performance than that directly trained on the same dataset.
For example, in \cite{Bucilua:2006:MC:1150402.1150464}, a fast and compact NN model is trained with pseudo data labeled by the ensemble of cumbersome NN models to approximate the function learned by the ensemble NNs.
It has been shown by the experiment results that, the NNs can be 1,000 times smaller and faster than the ensemble NNs with negligible performance loss.
Meanwhile, it can still alleviate the overfitting issue and has no memory and time costs of building an ensemble.

\subsection{Efficient NN Architecture Design}
\begin{figure}[t]
    \centering 
    \includegraphics[scale=0.5]{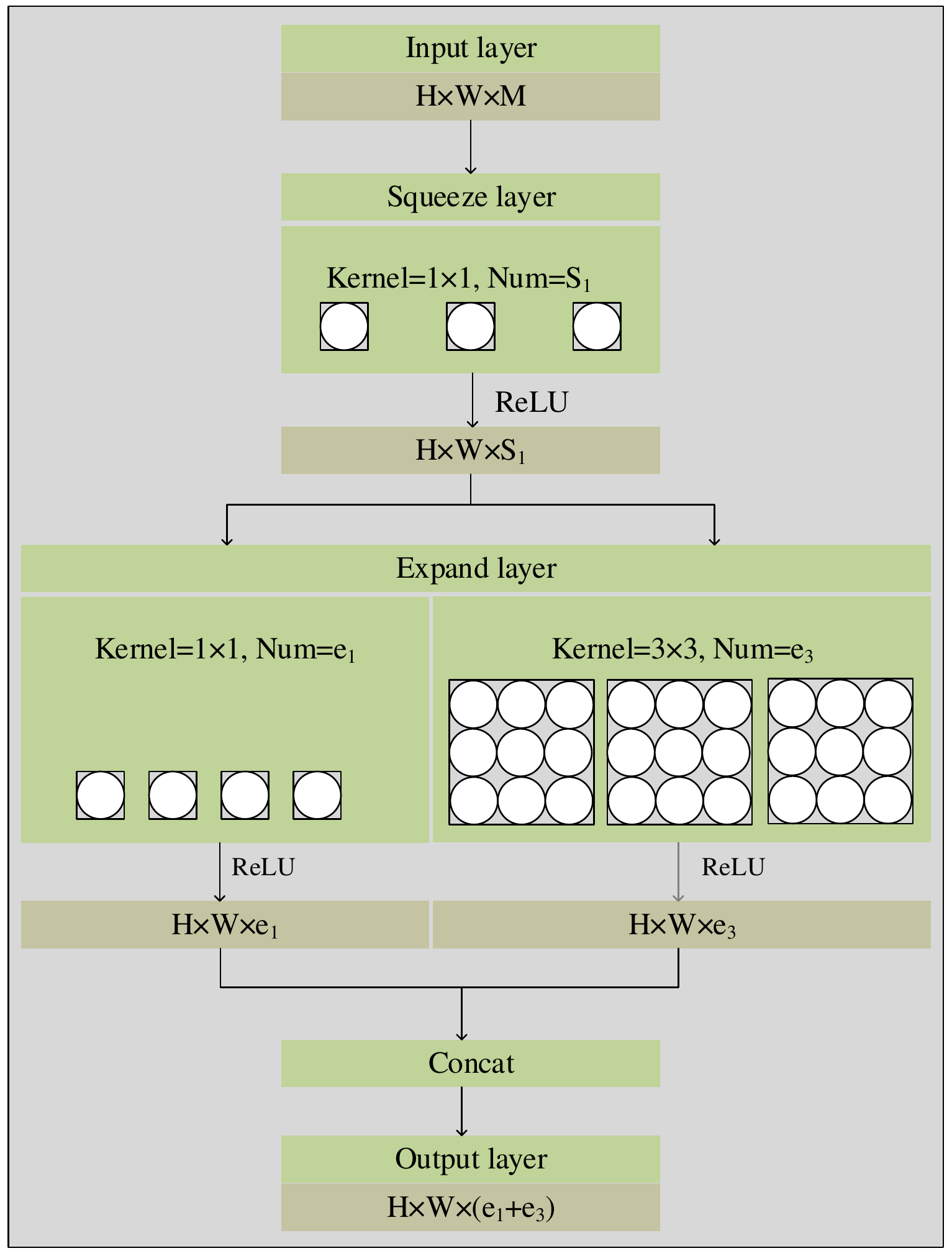}
    \caption{\label{squeezenet}Fire module architecture: a squeeze layer and a branching layer with $1\times1$ and $3\times3$ filters\cite{iandola2016squeezenet}.}  
\end{figure}

The goal of efficient NN architecture design is to make the NNs less redundant, i.e., decreasing the number of the network weights and computations straightforward with only limited model performance loss. 
FC layers are the most widely used NN layers in communications.
But they contain substantial weights and often lead to overfitting, thus hampering the NN generalization ability.
The global average pooling layers can be used to replace all FC layers to compress and accelerate NNs\cite{lin2013network}.
There are no weights to optimize in these layers, thus avoiding overfitting.
Also, they are more native to the feature extract modules by enforcing correspondences between classification labels and feature maps.
Therefore, the feature maps can be easily regarded as classification confidence maps.
Meanwhile, these layers sum out the spatial information and increase its robustness to spatial translations.

Convolution plays an important role for the NNs in communication systems.
Compared with FC layers, convolutional layers are better feature extractors, which take advantage of the local spatial coherence and have fewer weights via weight sharing strategy.
Convolutional layers have been used to CSI feedback \cite{guo2019convolutional}, channel decoding\cite{8259241}, etc.
When the channel\footnote{The `channel' here is totally different from that in communications.
Each channel is coressponding to a feature map. For examples, RGB images have 3 channels.} numbers of the input and the output feature maps, feature map size, and the convolutional kernel size are $C_{\mathrm{in}}$, $C_{\mathrm{out}}$, $H \times W$, and $K\times K$, respectively, the weight number and floating point operations (FLOPs) of a convolutional layer will be $C_{\mathrm{in}}\left(K^{2}+1\right) C_{\mathrm{out}}$ and $2 H W\left(C_{\text { in }} \times K^{2}+1\right) C_{\text { out }}$, respectively.
Even if they are much fewer than these in FC layers, decreasing the above hyperparameters is the key of the efficient but low-cost convolution operation design.
Therefore, $1\times 1$ filters are often first utilized to reduce the dimensionality of input features.
To decrease $K$, $1\times 1$ or $3\times 3$ filters rather than those with large sizes are stacked to extract features.

Another widely used strategy for reducing convolution complexity is group convolution, where the filters of a convolutional layer are split into multiple groups and the decreased weight number is proportional to the group number.
SqueezeNet\cite{iandola2016squeezenet} is one of the representative efficient CNNs.
The core block in SqueezeNet is the fire module, which consists of a squeeze layer and an expand layer and follows the aforementioned design principles, as in Fig. \ref{squeezenet}.
It achieves AlexNet-level classification accuracy but with 50$\times$ fewer NN weights.

\subsection{Network Pruning}
The performance of a NN can be improved by adding NN layers and neural neurons. 
Sometimes, a tiny performance improvement may incur a huge increase in the network depth and weight number, introducing substantial redundancy and complexity.
To remove these redundant connections and neurons that are unimportant and with less contribution to performance,  network pruning has been widely studied.
The basic idea of network pruning is to drop these weights with small absolute values.
This operation can introduce two benefits to NN compression and acceleration.
There are fewer weights needing to be stored, thereby saving the memory space.
Also, the computational operation involving these pruned weights are no longer needed, thereby reducing the computational complexity of NNs.

According to the granularity of pruning operation, the common pruning techniques can be divided into five groups: fine-grained, vector-level, kernel-level, group-level, and filter-level prunings.
The fine-grained pruning removes weights in an unstructured way, i.e., without considering weight locations.
This method leads to high sparsity of network weight but is not friendly to implementation since extra memory space is occupied to store indices that indicate the location of each pruned weight.
The vector-level and kernel-level pruning methods remove the dispensable vectors and 2-dimension (2D) kernels in the filters, respectively.
The group-level pruning method drops the weight at the same location of the filter.
In the filter-level method, the unimportant filters are pruned, which makes the NNs thinner.
The vector-level, kernel-level, and filter-level are friendly to hardware implementation since they prune weights in a structured way.

\subsection{Network Quantization} 
Network quantization includes the weight and the activation quantization\footnote{Gradient quantization, focusing on accelerating NN training, is also a model quantization method and we, however, just concentrate on those speeding up the inference stage.} and is another effective way to save memory space, speed up computations, and reduce memory access. 
In particular, weight quantization reduces the number of bits used to represent per weight and is the most widely used quantization technique.
Activation quantization replaces the substantial floating point multiply-accumulate operations in the activation layers with binary operations, thereby further speeding up the inference.

When quantizing the weights or activations, we can use the fixed or the adaptive codebook.
The fixed codebook quantization methodology is fixed-point quantization, where the codebook is predefined.
For example, in the binaried NNs, all network weights are quantized by $\rm Sign(x)$ function to $\{-1,1\}$.
The basic issue of this methodology is how to pre-define the codebook since it has great effects on the performance of quantized NNs.
In the adaptive codebook methodology, the codebook is learned from the weight dataset rather than predefined.
Therefore, the adaptive codebook quantization methodology can avoid the extra modifications to the training algorithms since NN weights are quantized after training.

Quantization can be performed deterministically or stochastically.
Rounding is perhaps the simpliest method of deterministic quantization; but the NN performance drops after this operation.
Vector quantization is also applied to NN quantization.
Its basic idea is to cluster the weights into several groups and then use the centroid of each group to represent the corresponding weights.
K-means algorithm is usually used to cluster weights.
It however is with expensive computation since the weight number is pretty large. 
Both of rounding and vector quantization ignore the features or distributions of the weights of NNs.
In the stochastic quantization, random rounding acts as a regularizer, injecting noises to NNs while the probabilistic quantization quantizes weights according to the weight distributions.

\subsection{Low-Rank Approximation}
The convolutional kernel $\mathbf{W}\in R^{w\times h \times c \times n}$ of the convolutional layers is a 4-D tensor,
where $w$, $h$, $c$, and $n$ denote the kernel width, kernel height, and the numbers of the channel of the input and output feature maps, respectively.
Reducing the redundancy in these 4-D tensors by merging some of the dimensions can greatly decrease the computational cost and memory requirement.
The basic issue here is to find an approximate tensor $\hat{\mathbf{W}}$ to represent the high-dimension tensor $\mathbf{W}$.
According to the number of components, this method can be divided into three kinds: 2-component, 3-component, and 4-component decomposition.
In the $n$-component decomposition, a fat convolutional layer is replaced with $n$ thin ones.
For example, for the 2-component decomposition, a $w \times h$ filter can be decomposed into two components: $w\times1$ and $1 \times h$ filters.
In other words, two convolutional layers, whose kernel sizes are $w\times1$ and $1 \times h$, respectively, replace the original $w \times h$ one, which not only reduces the weight number but also facilitates catching the horizontal and vertical correlations.

\subsection{Hardware Design}
The general-purpose platforms, e.g., powerful GPUs, cannot be deployed at the NN inference phase because of high monetary and energy cost.
Therefore, the specific platforms, which are computation-intensive and energy-efficient, should be designed.
Application Specific Integrated Circuit (ASIC) and Field-Programmable Gate Array (FPGA) are two promising hardware platforms\cite{8114708}.

ASIC is a kind of task-specific hardware and might be delicately designed to maxmize the benefits, e.g., power-efficiency and high throughput, in a specific NN implementation.  
The hardware parameters, however, are difficult to change once the DL-based algorithms are implemented on the ASIC.
Therefore, online training and NN model update are infeasible in the ASIC.
Different from ASIC, FPGA can be easily programmed and reconfigured and is friendly to online training and NN model update.
Meanwhile, the hierarchical storage structure and scheduling mechanism of FPGA can be optimized to improve the efficiency of accessing data, thereby reducing energy consumption.
  
The industry has invested a lot in the design of the novel NN accelerators.
For example, NVIDIA has released a highly flexible mobile multicore embedded System-on-Chip (SoC), namely, NVIDIA Tegra K1.
It has not only a high-performmance CPU cluster and GPU but also a low-performance and low-power CPU cluster.
The developer can fully control the operation setting to minimize the energy consumption.

\section{Two Case Studies for Massive MIMO}
\label{Two Case Studies}
We demonstrate the feasibility and potential of NN compression and acceleration techniques in the DL-based communications.
Since massive MIMO is a critical technique for future wireless networks, we present two case studies in massive MIMO systems: the CSI feedback based on an autoencoder architecture with substantial parameters and the signal detection based on FC layers with relatively few parameters.  
\begin{figure*}[t]
    \centering 
    \subfigure [NMSE performance of efficient network design.]{
     \includegraphics[scale=0.6]{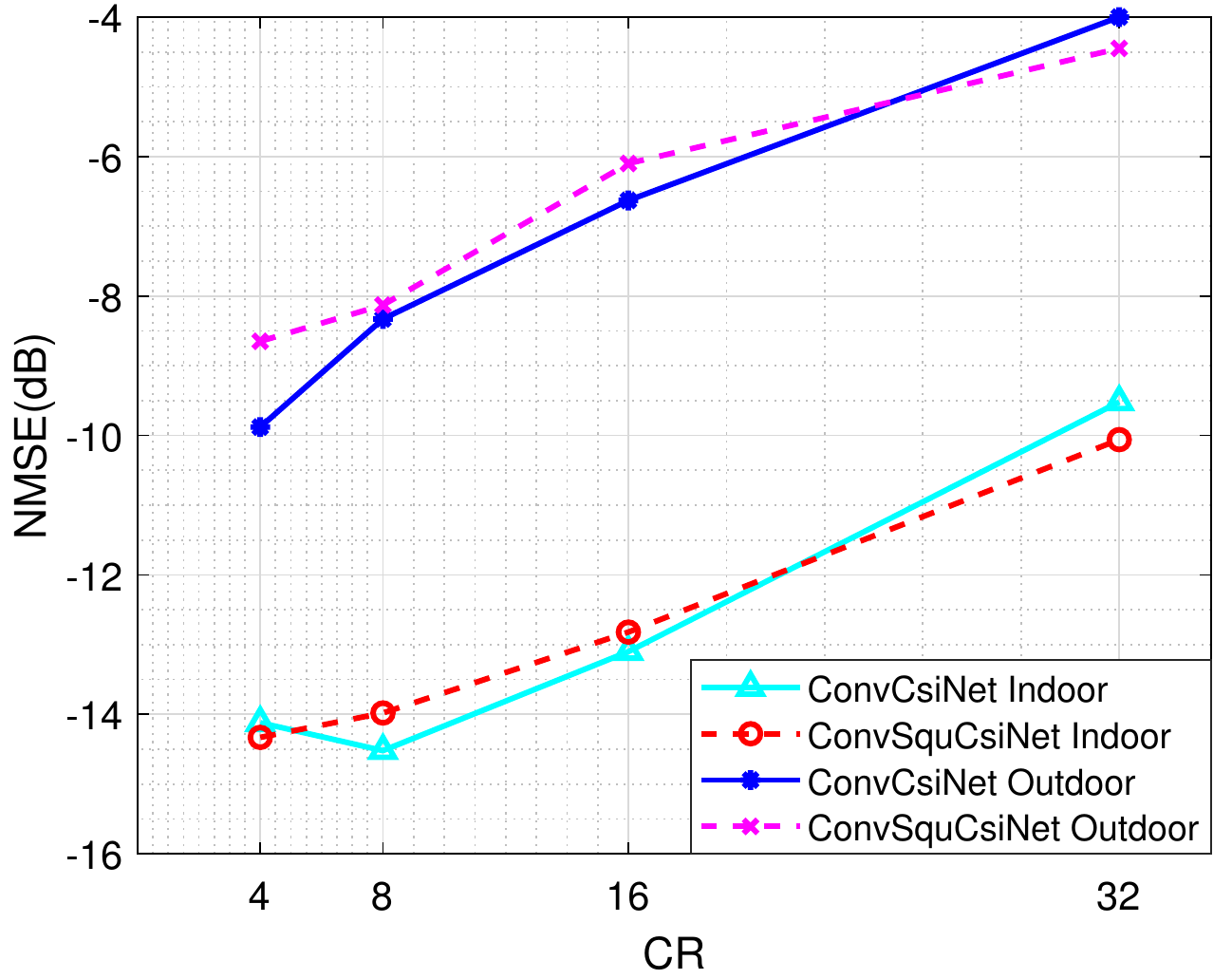}}   
     \subfigure [Weight number of efficient network design.]{
     \includegraphics[scale=0.6]{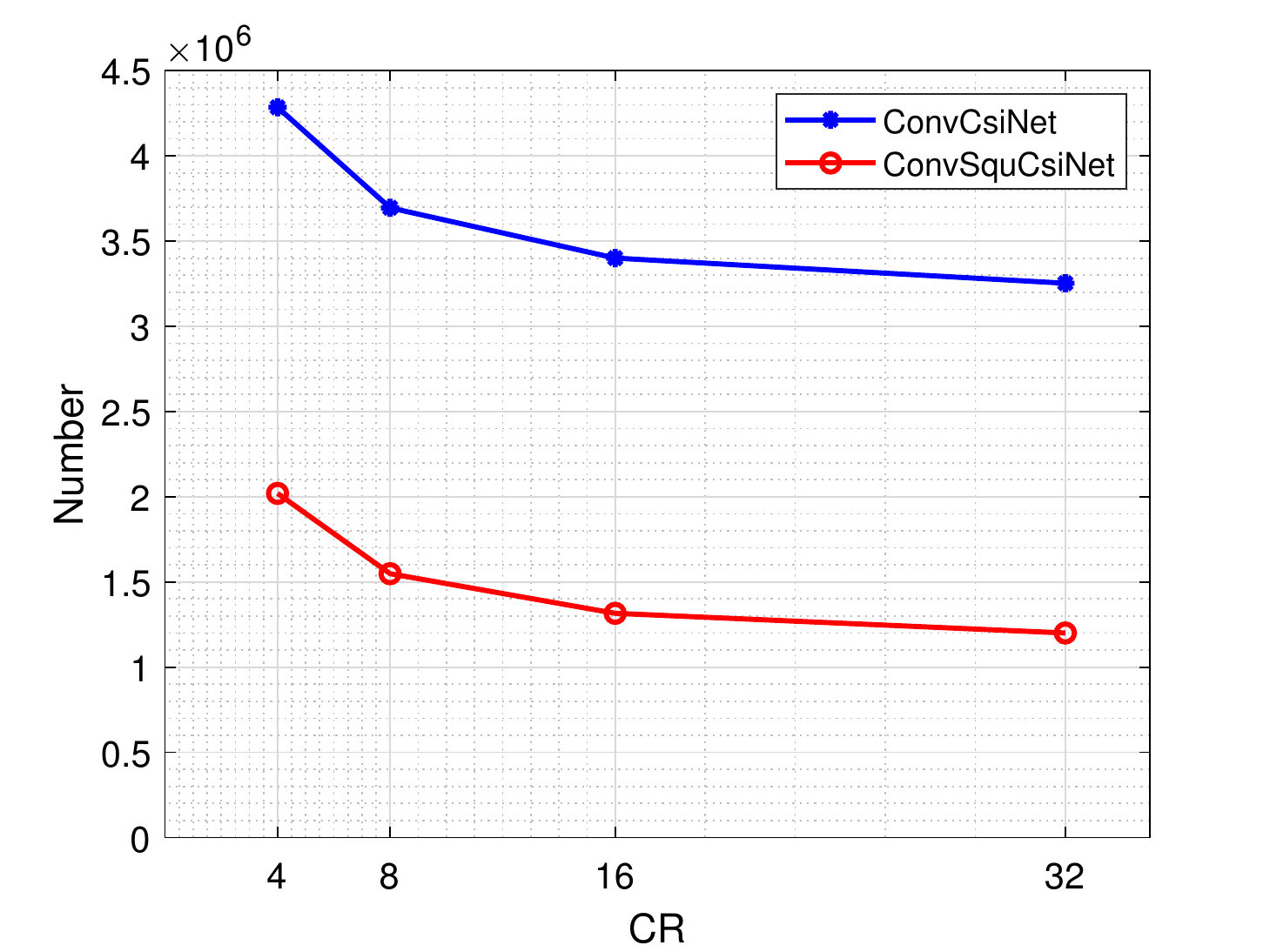}}           
	  \caption{\label{ConvSquCsiNet}NMSE performance and weight number comparison between ConvCsiNet and ConvSquCsiNet. } 
\end{figure*}

\subsection{CSI Feedback}
\label{csifeedback}
\begin{table*}[t]
\centering
\caption{\label{csinmse1}The NMSE ($dB$) of the pruned CsiNet+.}
\begin{threeparttable}

\begin{tabular}{c|c|cccc}
\hline   \hline
\multicolumn{2}{c|}{$CR$}                  & 4              & 8              & 16             & 32             \\  \hline
\multicolumn{2}{c|}{Original CsiNet+}      & -27.13         & -17.69         & -13.78         & -9.82                 \\  \hline
\multirow{5}{*}{ \rotatebox{90}{Indoor}}      & $t$=0.010 & -21.82(0.50\%) & -18.40(4.07\%) & -13.75(6.02\%) & -10.14(20.26\%)\\
                              & $t$=0.025 & -19.03(0.23\%) & -17.55(2.22\%) & -13.54(2.79\%) & -10.09(11.38\%) \\
                              & $t$=0.050 & -12.98(0.11\%) & -16.16(1.25\%) & -13.15(1.39\%) & -9.93(6.04\%)   \\
                              & $t$=0.075 & -9.63(0.07\%)  & -14.92(0.83\%) & -12.79(0.86\%) & -9.73(3.84\%)  \\
                              & $t$=0.100 & -8.49(0.06\%)  & -13.75(0.59\%) & -12.53(0.62\%) & -9.73(2.69\%)   \\ \hline\hline

\multirow{5}{*}{ \rotatebox{90}{Outdoor}}      & $t$=0.010 & -12.17(34.48\%) & -8.82(54.56\%) & -5.89(67.18\%) & -3.61(74.32\%) \\
                              & $t$=0.025 & -10.16(16.77\%) & -8.39(35.10\%) & -5.79(49.50\%) & -3.58(59.81\%) \\
                              & $t$=0.050  & -8.76(6.18\%)   & -6.66(17.60\%) & -5.39(32.32\%) & -3.44(44.55\%) \\
                              & $t$=0.075 &  -8.43(2.55\%)   & -5.10(8.37\%)  & -4.72(20.64\%) & -3.19(32.92\%) \\
                              & $t$=0.100    & -8.18(1.19\%)   & -5.05(3.81\%)  & -4.06(12.82\%) & -2.93(24.01\%) \\ \hline\hline                                             
 \end{tabular}
 \begin{tablenotes}
        \footnotesize
        \item[]Note: (.) denotes the remaining weight proportion after pruned.
      \end{tablenotes}
    \end{threeparttable}
         
\end{table*}
The benefit achieved by massive MIMO in communication systems is dependent on the accuracy of available CSI.
In frequency-division duplexing (FDD) systems, the UE has to constantly feed CSI back to the BSs for precoding.
With the increase of antenna number, the feedback overhead sharply increases, thereby leading to a large overhead and occupying precious bandwidth.
As a result, greatly compressing CSI before feeding it back is critical in massive MIMO systems.

The DL-based CSI feedback method \cite{wen2018deep} uses an encoder-decoder architecture to compress and reconstruct CSI at the UE and the BS, respectively, and outperforms the traditional compressive sensing (CS) algorithms by a margin.
With the two principles for DL-based CSI feedback network design, the CsiNet+ in \cite{guo2019convolutional} achieves much higher reconstruction accuracy than the original CsiNet \cite{wen2018deep} but only with a slight increase in parameter number.
To overcome the constraint of the fixed antenna number in the CsiNet caused by the FC layers, the ConvCsiNet in \cite{shih2018study} replaces the FC layers at the encoder and the decoder with stacked convolutional layers, each of which is followed by an average pooling layer and an upsampling layer, respectively.

%
%
%

In this case study, we will prune and quantize CsiNet+, respectively, and design an efficient NN architecture based on ConvCsiNet for CSI feedback.
We first train the CsiNet+ using an end-to-end approach.
Then, we prune the weights in the FC layers with a threshold $t$ and the NNs are retrained until convergent.
Afterwards, the weights in pre-trained NNs are quantized using k-means clustering and retrained until convergent.
The key idea of designing an efficient architecture based on ConvCsiNet is to reduce the dimension of convolutional kernels with the fire module in Fig. \ref{squeezenet}.
We call this modified ConvCsiNet as ConvSquCsiNet and train it from scratch.

As in \cite{wen2018deep,guo2019convolutional}, the datasets contain two representative scenarios, i.e., indoor and outdoor scenarios.
The pruning threshold $t$  for two FC layers is set as 0.010, 0.025, 0.050, 0.075, and 0.100, respectively.
The weights of CsiNet+ are quantized with 3-7 bits, respectively.
Normalized MSE (NMSE) is used to measure the CSI reconstruction accuracy.

\begin{table}[b]
\caption{\label{csinmse2}The NMSE ($dB$) of the quantized CsiNet+.}
\centering
\begin{tabular}{c|cccc|cccc}
\hline \hline
                 & \multicolumn{4}{c|}{Indoor}                                         & \multicolumn{4}{c}{Outdoor}                                        \\  \hline

$B$=32    & -27.13         & -17.69         & -13.78         & -9.82           & -11.36          & -8.28          & -5.60          & -3.42          \\
                               $B$=7     & -15.38         & -15.56         & -13.09         & -9.64           & -10.69          & -8.17          & -5.51          & -3.37          \\
                              $B$=6     & -11.60         & -13.21         & -11.51         & -9.02           & -9.81           & -7.73          & -5.21          & -3.11          \\
                              $B$=5     & -8.37          & -10.17         & -8.95          & -7.62           & -8.29           & -6.79          & -4.35          & -2.57          \\
                               $B$=4     & -3.91          & -6.37          & -5.88          & -5.27           & -5.97           & -4.29          & -2.83          & -1.60          \\
                               $B$=3     & -1.84          & -3.73          & -3.47          & -1.49           & -2.10           & -1.51          & -0.20          & -0.06         \\ \hline\hline
\end{tabular}
 \end{table}

Table \ref{csinmse1} shows the reconstruction accuracy versus compression rate ($CR$) with different pruning thresholds and the numbers of remaining weights after pruned.
Surprisingly, the pruned CsiNet+ even performs better than the original one when $CR$ = 16 or 32, $t$ = 0.010, 0.025 or 0.050, where more than 80\% and 30\% weights are pruned for the indoor and the outdoor scenarios, respectively.
Since there are too many redundant weight connections in the original FC layers, pruning operation can reduce the redundancy, improve the generality of CsiNet+, and help prevent overfitting.

\begin{figure*}[t]
    \centering 
    \subfigure [\label{case21}BER performance of pruned FullyCon.]{
     \includegraphics[scale=0.6]{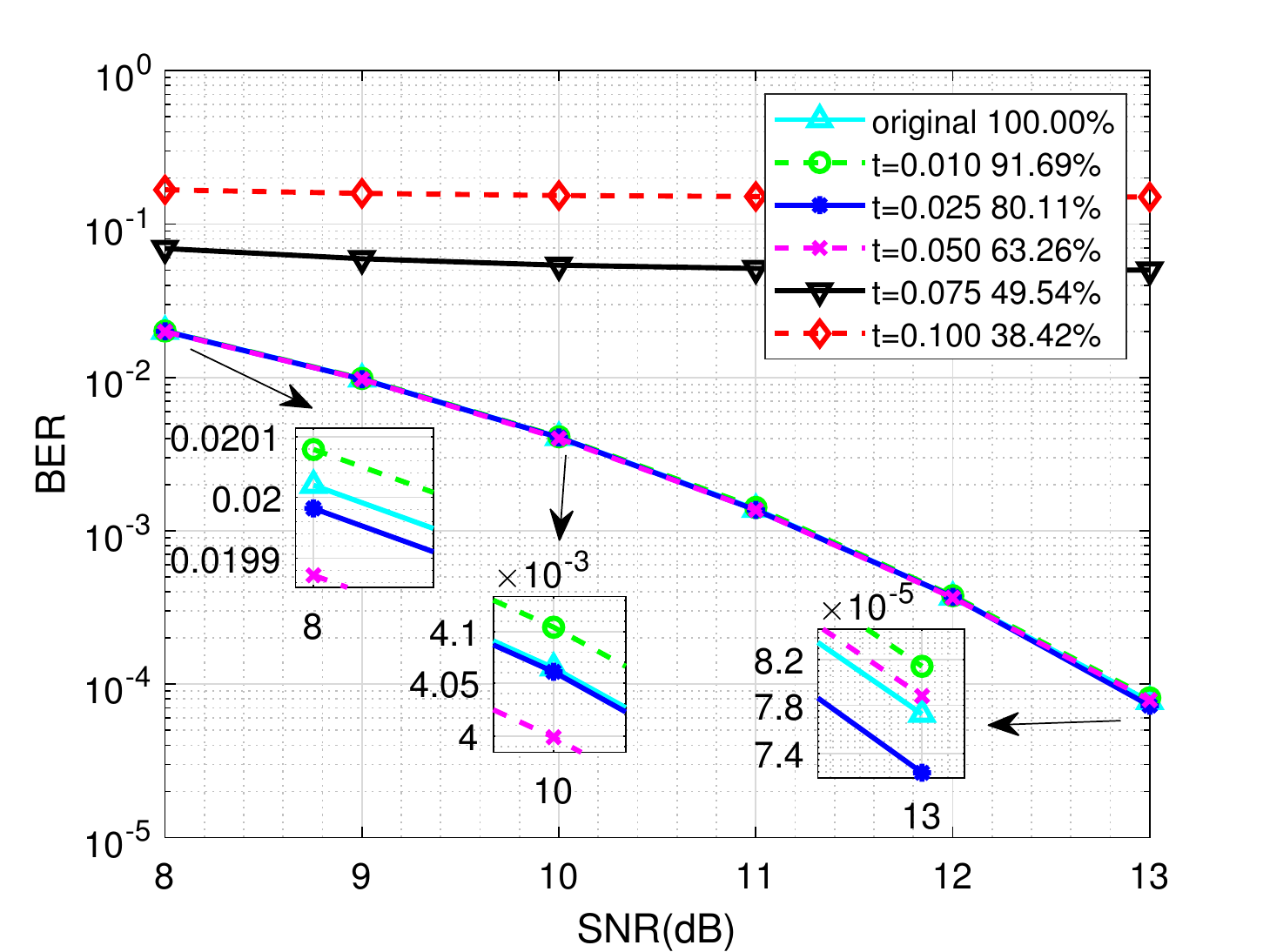}}   
     \subfigure [\label{case22}BER performance of quantized FullyCon.]{
     \includegraphics[scale=0.6]{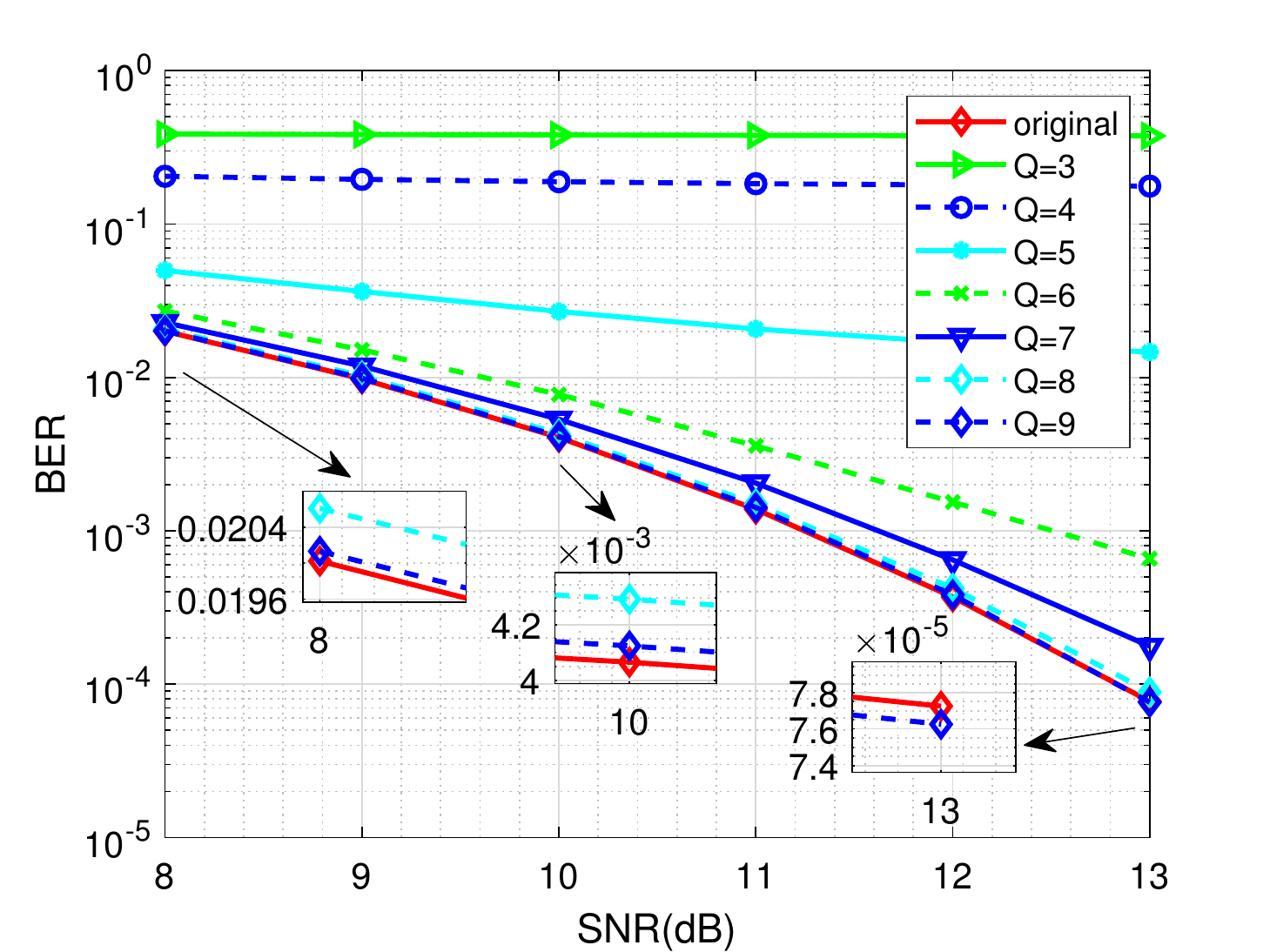}}           
	  \caption{\label{case2}BER performance comparison between FullyCon and the pruned or quantized one.} 
\end{figure*}

The CSI reconstruction accuracy of the quantized CsiNet+ corresponding to different quantization bits $B$ is shown in Table \ref{csinmse2}.
With the increase of quantization bits $B$, the performance of NNs are improved as we can imagine.
The quantized CsiNet+ can even have a similar accuracy as the original CsiNet+ without weight quantization when $CR$ = 16 or 32 and $B$ = 7.
Since the default quantization bits of network weights are set as 32-bit floating point, the memory space used to store network weights can reduce as more as 78\%, thereby greatly saving memory requirement and the power used to access data.

Fig. \ref{ConvSquCsiNet} shows the reconstruction performance and total weight number of ConvCsiNet and ConvSquCsiNet.
From that, ConvSquCsiNet has high reconstruction performance comparable to ConvCsiNet with fewer than half network weights and even outperforms ConvCsiNet when $CR$ = 32.

\subsection{Signal Detection}
\label{detection}

Inspired by the success achieved by DL in communications,
a FC layer-based NN, namely FullyCon, is introduced in \cite{samuel2019learning} to realize MIMO detection without an iterative operation.
The inference time of FullyCon in a fixed channel has an order of magnitude decrease compared with approximate message passing (AMP) algorithm.
The FullyCon contains about 211,220  NN weights, which can be pruned and quantized to further reduce the complexity.

The NN in FullyCon consists of FC layers.
The input layer has $N$ neurons, which are determined by the received signal size.
There are 4 FC layers with $10K$ neurons followed by the Rectified Linear Unit (ReLU) activation function, where $K$ is the symbol length.
The last layer has $K$ neurons to output the classification probability of each symbol.
We use 4 hidden layers in this case study instead of 6 hidden layers in \cite{samuel2019learning} since we focus on the effects of NN compression rather than improving the performance of signal detection.

In the signal detection, we only prune and quantize the NN weights, respectively.
Since all hidden layers in the FullyCon are FC layers, all weights are pruned and quantized.
The pruning threshold $t$  is set as 0.010, 0.025, 0.050, 0.075, and 0.100, respectively.
The weights are quantized using 3-9 bits, respectively.
We first train the FullyCon from scratch with a large learning rate and then prune and quantize all weights in the trained FullyCon, respectively.
Finally, the pruned and quantized FullyCon models are respectively retrained with a relatively small learning rate until converged. 

All experiments are performed on a fixed channel of size $30\times20$, which means that the signal length $N$ = 30 and the symbol length $K$ = 20.
The transmit symbols are modulated by BPSK.
The original learning rate is 0.001 and the one for pruning and quantization is 0.000,1.
The batch size is 1,000 and the FullyCon is optimized using Adam optimizer.
The SNRs of test scenarios are set as 8, 9 ,10 11, 12, and 13 dB, as in \cite{samuel2019learning}.

Fig. \ref{case21} shows the effects of network pruning on signal detection.
When the pruning threshold $t$ is 0.01, 0.025, and 0.05, i.e., 8.31\%, 19.89\%, and 36.74\% of the total weights are pruned, there is nearly no impact on the bit-error rate (BER).
With the increase of threshold $t$, the BER of pruned FullyCon will rise rapidly since the redundant connections have been dropped and the remaining are all dominant.
Therefore, finding a suitable pruning threshold is critical and should be carefully determined by extensive experiments.

In Fig. \ref{case22}, the BER drops with the increase of quantization bits.
When $B$ = 6, the BER of the quantized FullyCon rapidly rises.
If $B$ = 9, its performance is close to the original FullyCon without quantization operation.
In this scenario, about 71.878\% memory space is saved nearly without performance loss compared with 32-bit floating point.

\section{conclusion and discussions}
\label{conclusion}
In this article, we have investigated accelerating the deployment of DL-based algorithms in communications, which usually need large storage space and have high computational complexity.
We have introduced the NN compression and acceleration techniques to tackle the above challenges, including knowledge distillation, compact NN architecture design, network pruning, weight quantization, and low-rank approximation.
We have then demonstrated how to apply them to two representative problems in massive MIMO systems: CSI feedback and signal detection.

Encouraged by existing research results, there are still some issues needing to be addressed in NN compression and acceleration for future wireless communications.

(1) The performance loss is unavoidable when a NN model is compressed.
The tradeoff between the accuracy and the efficiency should be balanced according to specific hardware configurations and communication tasks.
For example, the UE has limited memory space and computational power, which however are not a constraint at the BSs.
Therefore, more attention should be paid to the compression at the UE.

(2) Different from the computer version problems that often contain just an NN model, communication systems may contain many DL-based modules.
If each DL-based module is compressed, the error caused by compression might accumulate, thereby greatly affecting the performance of the whole communication systems.
Hence, how to reduce this kind of errors should be taken into consideration. 
Meanwhile, different modules should be compressed to varying degrees since the performances of different modules have different effects on the final performance of the whole communication systems.

(3) For the most compression techniques, finding the optimal hyperparameters and the balance between accuracy and computational cost is time-consuming and requires substantial experiments.
Therefore, it will be great if developing an efficient way to determine the hyperparameters and balance.

(4) There is still a big room to exploit prior in compression because most of NN-based methods are fully data-driven.
However, model-driven DL \cite{8715338} is very promising in future communications.
In the compression of the communication NNs, expert knowledge should be full exploited rather than just using pure data-driven compression.

\bibliographystyle{IEEEtran}
\bibliography{IEEEabrv,magazine}

\begin{IEEEbiographynophoto}{Jiajia Guo}
(jjguo@mail.ustc.edu.cn) received the B.S. degree
from Nanjing University of Science and Technology, Nanjing, China, in 2016, and the M.S. degree from University of Science and Technology of China, Hefei, China, in 2019. He is currently
working towards the Ph.D. degree in information and communications engineering, Southeast
University, China, under the supervision of Prof. Shi Jin. His areas of interests currently include
deep learning for wireless communications and massive MIMO.
\end{IEEEbiographynophoto}

\begin{IEEEbiographynophoto}{Jinghe Wang}
(jwangeh@connect.ust.hk) received the B.S. degree
from Nanjing University of Science and Technology, Nanjing, China, in 2018, and the M.S. degree from the Hong Kong University of Science and Technology, Hong Kong, China, in 2019. She is currently
working towards the Ph.D. degree in information and communications engineering, Southeast
University, China, under the supervision of Prof. Shi Jin. Her areas of interests currently include
deep learning for wireless communications and massive MIMO.
\end{IEEEbiographynophoto}

\begin{IEEEbiographynophoto}{Chao-Kai Wen}
(chaokai.wen@mail.nsysu.edu.tw) received the Ph.D. degree from the Institute
of Communications Engineering, National Tsing Hua University, Taiwan, in 2004. He was with
Industrial Technology Research Institute, Hsinchu, Taiwan and MediaTek Inc., Hsinchu, Taiwan,
from 2004 to 2009. Since 2009, he has been with National Sun Yat-sen University, Taiwan,
where he is Professor of the Institute of Communications Engineering. His research interests
center around the optimization in wireless multimedia networks.
\end{IEEEbiographynophoto}

\begin{IEEEbiographynophoto}{Shi Jin}
[SM’17] (jinshi@seu.edu.cn) received the Ph.D. degree in communications and information systems from Southeast University, Nanjing, in 2007. From June 2007 to October 2009,
he was a Research Fellow with the Adastral Park Research Campus, University College London,
London, U.K. He is currently with the faculty of the National Mobile Communications Research
Laboratory, Southeast University. His research interests include space-time wireless communications, random matrix theory, and information theory. Dr. Jin and his coauthors received the 2010
Young Author Best Paper Award by the IEEE Signal Processing Society and the 2011 IEEE
Communications Society Stephen O. Rice Prize Paper Award in the field of communication
theory.
\end{IEEEbiographynophoto}

\begin{IEEEbiographynophoto}{Geoffrey Ye Li}
[F'06] (liye@ece.gatech.edu) is a Professor with Georgia Tech. His general research
is in signal processing and machine learning for wireless communications. In these areas, he
has published over 500 articles with over 36,000 citations and been listed as a Highly-Cited
Researcher by Thomson Reuters. He has been an IEEE Fellow since 2006. He won IEEE ComSoc
Stephen O. Rice Prize Paper Award and Award for Advances in Communication, IEEE VTS
James Evans Avant Garde Award and Jack Neubauer Memorial Award, IEEE SPS Donald G.
Fink Overview Paper Award, and Distinguished ECE Faculty Achievement Award from Georgia
Tech.

\end{IEEEbiographynophoto}

\ifCLASSOPTIONcaptionsoff
  \newpage
\fi

\end{document}